# Phonon density of states, anharmonicity, electron-phonon coupling and possible multigap superconductivity in the clathrate superconductors $Ba_8Si_{46}$ and $Ba_{24}Si_{100}$: Why is $T_c$ different in these two compounds?

Rolf Lortz[1†], Romain Viennois[1], Alexander Petrovic[1], Yuxing Wang[1], Pierre Toulemonde[2], Christoph Meingast[3], Michael Marek Koza[4], Hannu Mutka[4], Alexei Bossak[5], Alfonso San Miguel[6]

[1]Department of Condensed Matter Physics, University of Geneva, 24 Quai Ernest-Ansermet, CH-1211 Geneva 4, Switzerland
[2]Institut Néel - Département MCMF, CNRS and Université Joseph Fourier, 25 avenue des Martyrs, BP 166, F-38042 Grenoble cedex 9, France
[3]Institut für Festkörperphysik, Forschungszentrum Karlsruhe, Germany
[4]Institut Laue Langevin, 6 rue Jules Horowitz, F-38042 Grenoble cedex 9, France
[5]European Synchrotron Radiation Facility, 6 rue Jules Horowitz, F-38043 Grenoble cedex 9, France
[6]Université de Lyon, France; Univ. Lyon 1, Laboratoire PMCN; CNRS, UMR 5586; F-69622 Villeurbanne

**Abstract**
We report a detailed study of specific heat, electrical resistivity and thermal expansion in combination with inelastic neutron and inelastic X-ray scattering to investigate the origin of superconductivity in the two silicon clathrate superconductors $Ba_8Si_{46}$ and $Ba_{24}Si_{100}$. Both compounds have a similar structure based on encaged barium atoms in oversized silicon cages. However, the transition temperatures are rather different: 8 K and 1.5 K respectively. By extracting the superconducting properties, phonon density of states, electron-phonon coupling function and phonon anharmonicity from these measurements we discuss the important factors governing $T_c$ and explain the difference between the two compounds.

---

[†] Corresponding author. e-mail: Rolf.Lortz@physics.unige.ch

# I. Introduction

Novel materials based on structures with metallic ions located in oversized crystalline cages are an intriguing family. The encaged ions form a nanoscale crystalline subnetwork and influence a wide variety of physical properties. Amongst these materials those that become superconducting are particularly interesting since they represent model systems to study the electron-phonon interactions which mediate superconductivity. Examples of such systems include borides (e.g. $YB_6$ and $ZrB_{12}$)[1,2,3], b-pyrochlore superconductors $AOs_2O_6$ (A=Cs, Rb, K with $T_c$= 3.3, 6.3 and 9.6 K, respectively)[4,5], certain Chevrel-type molybdenum clusters[6] and, in principle, the prominent superconductor $MgB_2$ ($T_c$ = 39 K)[7,8], although in the latter case the cages are represented by open layers in between 2d boron planes. Similar systems are furthermore found in metallic cubic fullerides $A_3C_{60}$; however here the alkali (A) ions are intercalated between the cages[8].

The filled clathrates based on group IV elements (i.e. Si, Ge…) also belong to this family of materials and include several superconductors[9,10,11,12]. Type-I (e.g. $Ba_8Si_{46}$) and type-II (e.g. $Na_xSi_{136}$) Si-clathrates form three dimensional crystalline lattices based on rigid oversized 20 atom and 24 or 28 atom Si cages in which metal atoms are enclosed[13,14]. The type-III Si-clathrate (e.g. $Ba_{24}Si_{100}$) is built from the same "closed" $Si_{20}$ cages but also comprises "open" $Si_{20}$ cages and pseudo-cubic $Si_8$ cages[15].

In type-I and II Si clathrates, the metal ions are located in a cage-like crystalline host, formed by semiconducting $sp^3$-hybridized networks of Si. In the type-III $Ba_{24}Si_{100}$ clathrate, only 68 of the 100 Si atoms present a nearly pure $sp^3$ character. The study of such structures based on networks of covalent $sp^3$ bonds has recently attracted great attention following the discovery of superconductivity in diamond[16] and diamond-structured Si[17] which in both cases were doped with boron.

The mechanism of superconductivity was investigated both theoretically by *ab initio* calculations and experimentally[18,19] for the type-I $Ba_8Si_{46}$ clathrate. The joint experimental and theoretical study of Connétable *et al.*[18] has shown that superconductivity is an intrinsic property of the $sp^3$ silicon network. A large electron-phonon (e-ph) coupling (quantified by λ, the e-ph coupling constant) exists in such covalent structures. In addition, the $^{28}Si/^{30}Si$ isotope substitution and specific heat measurements of Tanigaki *et al.* proved that a BCS-type phonon-mediated superconductivity occurs in such compounds[19]. Furthermore, Toulemonde *et al.* have shown that the role of caged alkaline earth atoms in $M_8Si_{46}$ (M = Ba, Sr and Ca) is essentially to provide carriers to the network and enhance the electronic density of states at the Fermi level $N(E_F)$, without affecting significantly the e-ph coupling potential $V^{ep} = \lambda /N(E_F)$[20,21]. The effect on superconductivity of partially substituting the type-I Si host network by Ge, Ga, Au, Ag, etc. has also been studied experimentally over the last ten years: $T_c$ always decreases[22,23,24,25]. Recently, the theoretical work of Tsé *el al.* has pointed out that the low-frequency modes in $Ba_8Si_{46}$, in particular those arising from the Ba vibrations in the large $Si_{24}$ cages, contribute significantly to the electron-phonon coupling parameter λ[26]. In addition, their calculations also show that the low-frequency "rattling" modes of the Ba atoms (particularly the loosely-bound Ba in the large "open" $Si_{20}$ cages) are very efficient in mediating e-ph coupling for superconductivity in $Ba_{24}Si_{100}$[27].

In the boride[2,3] and pyrochlore superconductors[4,5], ions in similar oversized crystalline cages usually also exhibit low-energy vibrations ('Einstein phonons') which have been found to mediate superconductivity. The β-pyrochlore superconductors and the borides (e.g. $ZrB_{12}$ ($T_c$ = 6 K)[1] and $LuB_{12}$ ($T_c$ = 0.4 K)[28]) represent examples of families in which $T_c$ can be tuned by changing the mass of the caged ions. In the case of the two clathrate superconductors $Ba_8Si_{46}$ and $Ba_{24}Si_{100}$, it is the size of cages which varies, while the mass of the guest ion remains the same.

In this article we investigate the origin of superconductivity in $Ba_8Si_{46}$ and $Ba_{24}Si_{100}$. The two compounds have a similar structure but the transition temperatures are rather different: 8 K and 1.5 K, respectively. We extract the superconducting properties, phonon density of states (from specific heat measurements, inelastic neutron scattering and X-ray scattering), electron-phonon coupling function (electrical resistivity) and phonon anharmonicity (thermal expansion and inelastic neutron scattering). Our principal aim is to explain the difference of $T_c$ in the two compounds.

## II. Experimental details

$Ba_8Si_{46}$ and $Ba_{24}Si_{100}$ powder samples have been synthesized under high pressure and high temperature conditions[21]. In the $Ba_8Si_{46}$ sample, we observe diamond-Si impurities (about 1 %), whereas in the $Ba_{24}Si_{100}$ sample we observe diamond-Si (7 %) and cubic $BaSi_2$ (6 %) impurities. More details about the synthesis and structural characterization can be found in the references[11,20,21,29].

The specific heat was measured using a high-precision continuous-heating adiabatic calorimeter at high temperature between 14 and 300 K, then by a generalized long relaxation technique[30] at low temperature in a $^4$He cryostat between ~1.3 K and 15 K. Some measurements were repeated in a $^3$He cryostat. In this method, each relaxation provides about 1000 data points over a temperature interval of 30-40% of the base temperature, which has been varied between 1.3 K (400 mK for the $^3$He measurements) and 11 K.

The DC resistivity r was measured with a standard four probe technique from 2 to 300 K with current reversal using Degussa Leitsilber 200™ (a conducting silver paint) for the contacts. The residual resistivity of $Ba_8Si_{46}$ was obtained by suppressing the superconductivity with a magnetic field of 5.5 T. We obtained r(2K)=0.68 Ohm cm ($Ba_8Si_{46}$) and r(2K)= 24.4 Ω cm ($Ba_{24}Si_{100}$). As residual resistivity ratio we found r(300 K)/r(2 K)=1.77 ($Ba_8Si_{46}$) and 4.33 ($Ba_{24}Si_{100}$). The large resistivity values and low residual resistivity ratios arise from the polycrystalline nature of the samples.

A high-resolution capacitance dilatometer was used to measure the thermal expansion in the temperature range 5 – 300 K. Data were taken upon continuous heating at a rate of 15 mK/s with 20 mbar of $^4$He exchange gas used to thermally couple the sample to the dilatometer.

Inelastic neutron scattering (INS) experiments have been performed at the spectrometers IN4 and IN5 at the Institut Laue Langevin in Grenoble, France. The incident neutron wavelength was 2.25 Å with an energy resolution of 0.8 meV. Information on the GDOS has been obtained from the INS spectrum using the incoherent approximation[31]. These experiments were completed by inelastic X-ray scattering experiments (IXS) at the beam line ID28 of the European Synchrotron Radiation Facility in Grenoble, France.

### III. Electron specific heat and electron-phonon coupling strength

The specific heat of $Ba_8Si_{46}$ and $Ba_{24}Si_{100}$ at low temperature is presented in Figure 1. The superconducting state specific heat $C_S$ shows a sharp second-order jump at $T_c = 8.1$ K ($Ba_8Si_{46}$) and 1.55 K ($Ba_{24}Si_{100}$). Magnetic fields of 8 T ($Ba_8Si_{46}$) and 3 T ($Ba_{24}Si_{100}$) are sufficient to suppress superconductivity completely (see Figure 2). This allows us to analyze the normal-state specific heat in a standard way according to the expansion:

$$C_n(T \to 0) = \gamma_n T + \sum_{k=1}^{3} b_{2k+1} T^{2k+1},$$

where the first term is the electronic contribution, with $\gamma_n = \frac{1}{3}\pi^2 k_B^2 (1+\lambda_{ep}) N(E_F)$, $k_B$ Boltzmann's constant, $\lambda_{ep}$ the electron-phonon coupling constant and $N(E_F)$ the band-structure density of states at the Fermi level including two spin directions [i.e. the electronic density of states (EDOS)]. The second term is the low-temperature expansion of the lattice specific heat, where $b_3 = \frac{12}{5} N_{Av} k_B \pi^4 \theta_D^{-3}(0)$, with $N_{Av}$ Avogadro's number and $\theta_D(0)$ the initial Debye temperature. From a fit from 1.5 K to 5 K we obtain $\gamma_n = 2.29$ mJ/gat K$^2$ = 123.7 mJ/mole K$^2$ ($Ba_8Si_{46}$) and $\gamma_n = 1.53$ mJ/gat K$^2$ = 189.7 mJ/mole K$^2$ ($Ba_{24}Si_{100}$). The value for $Ba_8Si_{46}$ is slightly smaller than that reported in Ref.[19] (144 mJ/mole K$^2$). In a few of the samples we investigated, a residual $\gamma_{n\_res}$ arising from impurities was observed. Such a contribution may lead to a slight overestimation of $\gamma_n$. The $Ba_8Si_{46}$ sample used for the analysis in the present article exhibits a fully gapped nature at the lowest temperature without any residual $\gamma_{n\_res}$. The $T_c$ of $Ba_{24}Si_{100}$ is too low to clearly identify any residual $\gamma_{n\_res}$. However, our $\gamma_n$ value is close to that of Rachi et al.[12] ($\gamma_n = 182$ mJ/mole.K$^2$).

Figure 2 shows data for various magnetic fields for $Ba_8Si_{46}$ (a) and $Ba_{24}Si_{100}$ (c). The insets represent the phase diagram with the superconducting transition temperature as a function of the applied field. A fit using the standard Werthamer-Helfand-Hohenberg (WHH) theory[32] leads to estimates of the upper critical field $H_{c2} = 5.75$ T ($Ba_8Si_{46}$) and $H_{c2} = 0.35$ T ($Ba_{24}Si_{100}$). These values are in close agreement with estimates from other groups[11,12,33]. The dashed lines in Fig. 2a are extracted from the resistivity data (Fig. 2b) and correspond to the temperatures of the onset of the transition (stars), the midpoint (squares) and the temperature at which the resistance reaches zero (triangles).

Figure 3 displays in detail the electronic contributions to the specific heat of the two samples. The phonon contribution was separated using the low-temperature expansion of the specific heat as obtained from a fit to the normal state data. Also included is a fit using a standard BCS s-wave single-band α-model (dashed line). This yields an excellent fit above ~3 K, with a gap value of 0.47 meV corresponding to $2\Delta_0/k_B T_c = 4$. At lower temperature the data exceeds the fit, forming a small bump at ~1.5 K. This bump indicates an additional contribution from low-energy excitations which cannot be described by standard BCS behavior. Taking into consideration the scanning tunneling spectroscopy data from K. Ichimura et al.[34], who found evidence for an anisotropic s-wave gap with a minimum gap value of $2\Delta_{min}/k_BT_c = 1.3$ and a maximum $2\Delta_{max}/k_BT_c = 4.4$, we have performed fits with both an anisotropic s-wave gap (with an elliptical gap distribution) and a two gap model. We are able to approximate the characteristic features satisfactorily using the two gap model with gap values of $2\Delta_{0\_1}/k_BT_c = 1.0$ (10 %) and $2\Delta_{0\_2}/k_BT_c = 4.2$ (90 %). The contribution from the second gap appears to be rather small, comparable to what has been observed in $Nb_3Sn$[35]. It is certainly much smaller than in $MgB_2$[30], where the two gaps contribute equally to the specific heat. A d-wave order parameter can be ruled out: the specific heat does not show a linear temperature dependence at the lowest temperature and the jump at $T_c$ is too large. The gap values we find using the two gap model are close to the extreme gap values of Ichimura et al.[34]. However, if we try to fit our data with an anisotropic s-wave model instead[30] using the same extreme gap values, the fit clearly fails. A model with a continuous gap distribution cannot describe such a pronounced bump in the temperature dependence of the specific heat. In scanning tunneling data at a fixed temperature, it may however be difficult to distinguish a two-gap order parameter from an anisotropic s-wave gap[36], while the temperature dependence of the specific heat clearly favors a two-gap scenario. Recent band structure calculations[27] reveal that several bands cross the Fermi level in $Ba_8Si_{46}$. A multigap superconductivity is hence possible, but due to the complex band structure it is difficult to judge which of the bands are related to the two superconducting gaps.

In the case of $Ba_{24}Si_{100}$ (shown in Figure 3b), the best possible fit we can obtain with a BCS model is found for $2\Delta_0/k_BT_c = 3.0$. However, the fit shows a strong curvature due to the small gap value and lies above the data. This deviation may be due to some impurity residual $\gamma_{n\_res}$ term, similar to that observed in some of our $Ba_8Si_{46}$ samples. Due to our limited temperature range we were not able to conclusively identify this in $Ba_{24}Si_{100}$. If however we assume the BCS value for the relation $\Delta C/\gamma_n T_c = 1.43$ we obtain a corrected $\gamma_{n\_corr} = 1.28$ mJ/gat $K^2$ = 158.7 mJ/mole $K^2$. Subtraction of the estimated $\gamma_{n\_res}$ value dramatically improves the quality of the BCS fit with a gap value of $2\Delta_0/k_BT_c = 3.4$.

The Sommerfeld constants of both compounds correspond to a renormalized density of states at the Fermi level $(1+\lambda_{ep})N(E_F) = 52.5$ states/(eV cell) = 6.6 states/(eV Ba-atom) = 0.98 states/(eV atom) ($Ba_8Si_{46}$) and $(1+\lambda_{ep})N(E_F) = 80.4$ states/(eV cell) = 3.35 states/(eV Ba-atom) = 0.64 states/(eV atom) ($Ba_{24}Si_{100}$). If we take our corrected $\gamma_{n\_corr}$ for $Ba_{24}Si_{100}$ instead the latter value is reduced to $(1+\lambda_{ep})N(E_F) = 63.9$ states/(eV cell) = 2.66 states/(eV Ba-atom) = 0.51 states/(eV atom). The units per Ba-atom and per atom allow us to compare values for the two compounds: the density of states at the Fermi level in $Ba_{24}Si_{100}$ is clearly lower than in $Ba_8Si_{46}$, which may already partly explain its lower transition temperature.

The normalized specific-heat jump in $Ba_8Si_{46}$ is $\Delta C/\gamma_n T_c = 1.71$, which is in the medium coupling regime, above the weak-coupling BCS limit 1.43. This supports estimates of $0.8 < \lambda_{ep} < 1.2$ found in the literature[11,19,26]. Band-structure calculations show a pronounced peak at $E_F$ with a maximum of either 43 states/(eV cell)[18,37] or 38 states/(eV cell)[25]. In Ref.[18], $E_F$ is located at the peak while in Ref.[37] it is slightly above the peak. If we assume $1+\lambda_{ep} \cong 2$, our value of $N(E_F) = 26.25$ states/(eV cell) (which is even smaller than the value $N(E_F) = 31$ states/(eV cell) of Tanigaki et al.[19]), supports the scenario of Moriguchi et al.[37] in which the peak is located slightly below $E_F$. For $Ba_{24}Si_{100}$ a value as small as

$\Delta C/\gamma_n T_c$=1.1 is found, which is below the weak-coupling BCS limit. This value may be underestimated due to the presence of a possible residual $\gamma_n$ component, as we discussed earlier. Zerec et al.[38] report band structure calculations $N(E_F) \cong 60$ states/(eV cell). If we compare our value $(1+\lambda_{ep})N(E_F)$ = 80.4 states/(eV cell) (corrected value: 63.9 states/(eV cell)) we only find room for a small renormalization factor $1+\lambda_{ep}$= 1.34 (1.1) indicative of weak coupling. Table I gives an overview of the superconducting properties of $Ba_8Si_{46}$ and $Ba_{24}Si_{100}$.

## IV. Lattice specific heat and phonon density of states in comparison with microscopic dynamics and scattering experiments

The specific heat in the normal state of both $Ba_8Si_{46}$ and $Ba_{24}Si_{100}$ shows a rather unusual temperature dependence at low temperatures (Fig. 4). The $T^3$ regime of the lattice specific heat does not extend beyond a few Kelvin, as shown by the rapid increase of the specific heat at low temperatures in Fig. 1. This indicates the presence of peaks in the phonon density of states (PDOS) at low energies. To understand the origin of superconductivity and the difference in $T_c$ between the two compounds, it is instructive to investigate the PDOS in detail. We use several different methods: a deconvolution of the normal state specific-heat data into a set of Einstein modes[2,3], inelastic neutron scattering (INS) and inelastic X-ray scattering (IXS). Although the specific heat only provides a very limited energy resolution for the PDOS, this approach is useful for a later comparison with a similar deconvolution method in order to obtain the electron-phonon coupling constant from the resistivity and the Grüneisen parameter from the thermal expansion.

The specific-heat data at high temperature are sufficiently minimally scattered to attempt a deconvolution of $C_{ph}(T)$ to extract the PDOS $F(\omega)$. A simplified method consists of representing $F(\omega)$ by a basis of Einstein modes with constant spacing on a logarithmic frequency axis:

$$F(\omega) = \sum_k F_k \delta(\omega - \omega_k). \qquad (1)$$

The corresponding lattice specific heat is given by:

$$C_{ph}(T) = 3N_{Av}k_B \sum_k F_k \frac{x_k^2 e^{x_k}}{(e^{x_k}-1)^2} \qquad (2)$$

where $x_k = \omega_k/T$. The weights $F_k$ are found by a least-squares fit of the lattice specific heat. The number of modes is chosen to be small enough to ensure the stability of the solution. Note that we do not try to find the energy of each mode; we rather aim to establish a histogram of the density in predefined frequency bins. The robustness of the fit is demonstrated by the r.m.s deviation: <0.6% above 10 K.

Figure 4 illustrates the decomposition of the lattice specific heat into Einstein contributions. The PDOS obtained in this way is shown in Fig. 5 below. Several strong peaks at ~7 meV, ~17 meV and ~42 meV are found. For $Ba_{24}Si_{100}$ an additional peak appears at low energy at ~3 meV.

Although specific heat, used here as a "thermal spectroscopy", only has a limited energy resolution, it is sufficient to resolve the main peaks in the PDOS. However, care must be taken that the phonon modes do not shift significantly over the measured temperature range and we will show in the following that this assumption is justified. An advantage of the PDOS obtained from the specific heat by this fitting procedure is that the absolute values of the mode amplitudes may be determined in certain cases. This is due to the constraint that the sum over all Einstein components has to approach the value 3$R$ at high temperatures (where $R$ is the universal gas constant).

In order to clarify the origin of the peaks and justify our specific-heat approach, a comparison with standard determinations of the PDOS is instructive. The spectral distribution function $F(\omega)$ can be extracted from INS and IXS scattering experiments. In principle, both scattering techniques sample the scattering function $S(Q,w)$ reflecting the spatial and temporal correlation of atoms in a material[31,39]. Their space and time resolution accurately matches interatomic distances and vibrational dynamics, respectively. Thermal and cold neutrons are the best suitable experimental tools to illuminate the response of $F(\omega)$ to temperature changes with an accuracy of only a few 100 meV, i.e. a few Kelvin. The INS experiments have been performed at the spectrometers IN4 and IN5 at the Institut Laue Langevin in Grenoble, France. Experimental details can be obtained from reference[40]. In addition, X-rays can be utilized in a complementary fashion to probe the partial distribution of modes associated with the either the barium vibrations or the silicon dynamics. IXS experiments have been performed at the ID28 beamline of the European Synchrotron Radiation Facility in Grenoble, France. An energy resolution of 3 meV (full width at half maximum) and ambient temperature conditions were sufficient to extract the required information.

The complementarity of the two techniques results from the different coupling mechanisms of the two probes (neutrons and X-rays) with the sample material. The scattering probability of X-rays by the electronic shell of the sample constituents is Q-dependent and at low Q simply proportional to $Z^2$, the atomic number of the scatterer squared. It is therefore strongly barium biased. Neutrons are scattered from nuclei with a Fermi pseudo-potential governing the interaction strength. Table II indicates the relative coupling strengths of neutrons and X-rays with the sample constituent elements. As far as translational dynamics are concerned, the partial intensities are determined by the eigenmode amplitudes and are hence proportional to $1/M$ where $M$ is the mass of the scatterers. The resulting effective scattering intensity of the different elements is also indicated in Table II. Consequently, and without taking the stoichiometry of the compounds into account, we may state that INS is roughly three times more sensitive to silicon than barium modes and IXS around three times more sensitive to barium than silicon modes.

However, since we have been dealing with purely or predominantly coherent scatteres for X-rays and neutrons respectively our data interpretation of the data is influenced by the experimentally-accessible momentum range $Q$ and the cross-correlation terms between barium and silicon. Our data permits us to calculate the generalized density of states

$G(\omega)$ which accurately represents the positions of characteristic modes, but yields a less precise estimate of the partial intensity distribution of the eigenmodes[31,41].

Figure 6 displays the $G(\omega)$ of $Ba_8Si_{46}$ studied by INS and IXS and a $Ba_{24}Si_{100}$ sample measured by INS. From the relative intensity difference of the INS and IXS signals we may conclude that the barium eigenmodes in $Ba_8Si_{46}$ are located at energies below 25 meV. In particular the peak in the range 6 - 10 meV shows a substantial intensity change upon changing from X-rays to neutrons indicating a majority contribution from barium vibrations. This observation is in good agreement with previous INS measurements coupled to X-ray absorption spectroscopy measurements and *ab initio* calculations[42] as well as Raman-scattering and lattice dynamics results[26]. Considering whether a significant contribution from barium can also be found beyond the first strong peak, i.e. above 12 meV, we examine the effective number of modes in the INS and IXS experiments for the two separate intensity bands below and above 12 meV. This calculation overestimates the experimentally-determined intensity of the high energy band by a factor of two when compared with the model results. Taking account of the perturbations by monitoring the generalized density of states, we may conclude that a definite barium contribution has to be taken into account at energies above 12 meV. The second distinct peak in the IXS data at 15 meV (which is also visible as a shoulder in the INS spectra) arises from hybridized Ba - Si modes. Furthermore, the maximum in the INS data can be attributed to silicon modes as it is supported by Raman-scattering results[26]. This is also the case for the peak at 45 meV. A weak feature in the IXS data (not shown here) around 55 meV resembles results from lattice dynamics and Raman experiments; however, it cannot be confirmed within the INS experiments[40].

In the case of $Ba_{24}Si_{100}$ the INS measurement reveals a remarkably textured low-energy density of states, a feature which is corroborated by lattice dynamics calculations[27]. When following the effective mode argument outlined above, the experimentally determined intensities match the hypothetically expected intensities rather well. For $Ba_{24}Si_{100}$ the barium modes appear to be concentrated below 12 meV, in good agreement with the lattice dynamics results. This agreement also applies for the uniform distribution of silicon modes at higher energies and the very low energy barium mode around 2.5 meV, whose presence has been confirmed with a high resolution experiment at IN5.[40]

Figure 7 reports the temperature dependence of the clathrate samples measured by INS. The experimental data are presented as the generalized susceptibility, whose profile is independent of the temperature for a harmonic system since it has been corrected for the thermal occupancy of the vibrational modes. This presentation is favorable for stressing low-energy features, i.e. the regime in which the barium contribution dominates. For consistency with the energy scale of the generalized density of states, the anti-Stokes line (energy gain of the neutrons) is plotted on the positive scale.

Within the accuracy of the experimental approach and the temperature ranges studied we cannot identify significant changes in the spectral distributions of the two samples. The positions of distinct modes are unaltered upon temperature changes. Although subtle variations of the spectral intensity can be observed, their origin cannot be assigned to unequivocal changes of the sample signal. The rather weak scattering power of the barium modes in the INS experiments counterbalances the advantage of the excellent energy resolution at the lowest temperatures where background signal corrections become increasingly important. Nevertheless, the important point to be stressed is that the present data enable us to conclude that a temperature independent distribution of vibrational eigenmodes is an appropriate approximation within the studied temperature range of 2-300 K. This is an important result as a strong temperature dependence would invalidate our modeling of the PDOS from the specific heat, which probes the phonon energy scale as a function of temperature.

The question arises as to what extent the peaks in the PDOS contribute to the electron-phonon coupling. This point is addressed in the next section, using the sample electrical resistivity as an experimental probe.

## V. Resistivity and electron-phonon coupling

The brittleness and polycrystallinity of our samples poses in general an obstacle for the determination of the resistivity on an absolute scale. Not only is an excess value observed in polycrystalline samples compared to the bulk properties, but a variation can also be seen depending on the consistency of the sample material. We therefore do not attempt to extract any information from the absolute value of the resistivity, but focus on its temperature dependence which shows a characteristic form reproduced in all our samples over an extensive series of measurements. We analyze the resistivity (Figure 8) in a similar way to the specific heat. We start from the generalized Bloch-Grüneisen formula (see e.g. Ref.[43], in particular p. 212 and 219):

$$\rho_{BG}(T) = \rho(0) + \frac{4\pi m^*}{ne^2} \int_0^{\omega_{max}} \alpha_{tr}^2 F(\omega) \frac{xe^x}{(e^x-1)^2} d\omega \qquad (7)$$

where $x \equiv \omega/T$ and $\alpha_{tr}^2 F(\omega)$ is the electron-phonon "transport coupling function". In the restricted Bloch-Grüneisen approach, one would have $\alpha_{tr}^2 F(\omega) \propto \omega^4$ and as a consequence $\rho_{BG}(T) - \rho(0) \propto T^5$, but deviations from the Debye model, complications with phonon polarizations and Umklapp processes would not justify this simplification beyond the low-temperature continuum limit, i.e. only a few Kelvin in this case. Using a decomposition into a basis of Einstein modes similar to Eq. (1),

$$\alpha_{tr}^2 F(\omega) = \frac{1}{2} \sum_k \lambda_{tr,k} \omega_k \delta(\omega - \omega_k) , \qquad (8)$$

we obtain the discrete version of Eq. (9):

$$\rho_{BG}(T) = \rho(0) + \frac{2\pi}{\varepsilon_0 \Omega_p^2} \sum_k \lambda_{tr,k} \omega_k \frac{x_k e^{x_k}}{(e^{x_k}-1)^2} \qquad (9)$$

where the fitting parameters are the dimensionless constants $\lambda_{tr,k}$. The constraint $\lambda_{tr,k} \geq 0$ is enforced. The residual resistivity $\rho(0) = 0.68$ Ω cm (Ba$_8$Si$_{46}$) and $\rho(0) = 24.4$ Ω cm (Ba$_{24}$Si$_{100}$) is determined separately. $\Omega_p \equiv (ne^2/\varepsilon_0 m^*)^{1/2}$ is the unscreened plasma frequency. The negative curvature of the resistivity of Ba$_8$Si$_{46}$ at high temperature, a rather general phenomenon possibly related to the Mott limit[44,45], is taken into account by the empirical "parallel resistor" formula[46]:

$$\frac{1}{\rho(T)} = \frac{1}{\rho_{BG}(T) + \rho(0)} + \frac{1}{\rho_{max}} \,. \tag{10}$$

The parameters $\rho_{max} = 1.06$ Ω cm (Ba$_8$Si$_{46}$) and 103.6 Ω cm (Ba$_{24}$Si$_{100}$) are fitted simultaneously with the parameters $\lambda_{tr,k}$.

The results of the fits are shown in Figure 9 as $\alpha_{tr}^2 F(\omega)$ in comparison with the phonon density of states $F(\omega)$ as obtained from specific heat. The deconvolution of the resistivity yields a good reproduction of all the phonon peaks obtained from the deconvolution of the specific heat. Essentially, two modes at 7 meV and 17 meV are found for Ba$_8$Si$_{46}$, which represent the two peaks at lower energy in the PDOS. For Ba$_{24}$Si$_{100}$ all the peaks in the PDOS are reproduced. The strongest peak of the low energy modes appears at the additional peak in the PDOS at 3 meV which was confirmed by the high-resolution INS experiment at 2.5 meV. Due to the pronounced convex behavior in this compound we limited the fitting interval to 200 K. Otherwise, a huge peak appears at 42 meV, which is an artifact due to the parallel resistor model. This does not however affect the modes at lower energies.

The electron-phonon coupling parameter relevant for transport $\lambda_{tr} \equiv 2\int \omega^{-1} \alpha_{tr}^2 F(\omega)$ is related to the fitting parameters $\lambda_{tr,k}$ by $\lambda_{tr} = \sum_k \lambda_{tr,k}$. Due to the polycrystalline nature of our samples we have no access to the absolute values of the resistivity. We are therefore unable to extract this value and only use the relative weights $\lambda_{tr,k}$ in arbitrary units as a measure of the electron-phonon coupling strength.

When compared with thermodynamic data, the analysis of the DC conductivity leads to the conclusion that different phonon modes (or narrow groups of modes) are responsible for the superconductivity in Ba$_8$Si$_{46}$ and Ba$_{24}$Si$_{100}$. In Figure 9c we extract $\alpha_{tr}^2$ by dividing the spectrum obtained from the resistivity by $F(\omega)$. This graph clearly shows that the main contribution to the electron-phonon coupling in Ba$_{24}$Si$_{100}$ arises from the low energy mode in the PDOS at ~3 meV (2.5 meV from high-resolution INS). In Ba$_8$Si$_{46}$ the superconductivity is essentially driven by the mode at ~7 meV. This may partly explain the lower $T_c$ in Ba$_{24}$Si$_{100}$.

## VI. Thermal expansivity and anharmonicity

Thermal-expansion experiments were performed to give three types of information: (i) confirmation of the main features of the PDOS; (ii) evaluation of the volume dependence of the phonon modes and electronic density of states; (iii) determination of the variation of $T_c$ with pressure. Figure 10 shows the linear thermal expansion coefficient for $Ba_8Si_{46}$ and $Ba_{24}Si_{100}$. A drop occurs at $T_c$ of $Ba_8Si_{46}$ (see inset). The superconducting transition of $Ba_{24}Si_{100}$ is below the low-temperature limit of our equipment.

The linear thermal expansivity $\alpha(T)$ for a cubic system is given by:

$$\alpha(T) \equiv \frac{1}{L}\left(\frac{\partial L}{\partial T}\right)_p = \frac{\kappa_T}{3}\left(\frac{\partial S}{\partial V}\right)_T, \tag{11}$$

where $\kappa_T$ is the isothermal compressibility. The expansivity is closely related to the specific heat at constant volume via the Grüneisen parameters (see e.g. Ref.[43]):

$$\alpha(T) = \frac{\kappa_T}{3V}\left(\gamma_{G,el} C_{el} + \gamma_{G,ph} C_{ph}\right) \tag{12}$$

where the electronic Grüneisen parameter $\gamma_{G,el} = \partial \ln \gamma_n / \partial \ln V$ provides a measure of the volume dependence of the Sommerfeld constant and the phonon Grüneisen parameter $\gamma_{G,ph} \equiv -\partial \ln \omega / \partial \ln V$ represents the anharmonicity of the lattice vibrations. In this simple form, we can make use of the known electronic component $C_{el}(T)$ of the specific heat in the normal state and extract $\gamma_{G,el}$ from a fit to the normal-state expansivity curves $\alpha(T)$ at low temperature. As in the case of the specific heat, a plot of $\alpha/T$ versus $T^2$ (Inset of Figure 10) is most suitable for displaying the results. The data shows a linear behavior below 200 K$^2$ which we extrapolate to $T = 0$ to estimate the electronic component of the expansivity; $\alpha_{el}(T)/T = -8.4 \times 10^{-9}$ K$^{-2}$ ($Ba_8Si_{46}$) and $a_{el}(T)/T = +3.2 \times 10^{-9}$ K$^{-2}$ ($Ba_{24}Si_{100}$). These values are stable when the upper limit of the fit is varied between 100 and 200 K$^2$. Using the bulk modulus $\kappa_T^{-1} = 93$ GPa of $Ba_8Si_{46}$[47] and $\kappa_T^{-1} = 90$ GPa of $Ba_{24}Si_{100}$[48] allows us to calculate $\gamma_{G,el} = -12.6$ ($Ba_8Si_{46}$) and $\gamma_{G,el} = +7.6$ ($Ba_{24}Si_{100}$).

When calculating the phonon Grüneisen parameter $\gamma_{G,ph} \equiv -\partial \ln \omega / \partial \ln V$, we must take its frequency dependence into account. Modes which are characterized by a large $\gamma_{G,ph}(\omega)$ are more heavily weighted in the thermal expansion than in the specific heat. This is exemplified by the expansivity data shown in Figure 10. The expansivity of $Ba_8Si_{46}$ shows a stronger increase than that of $Ba_{24}Si_{100}$ at low temperatures, which is evidence for a larger volume dependence in some low-frequency modes. In order to evaluate the energy of these modes and compare the two different clathrate compounds, we fit the phonon expansivity over the full temperature range in a similar manner to the resistivity and the specific heat, using the same set of Einstein frequencies (Figure 11).

Equation (13) below, similar to Eq. (2) and (7), allows the parameters $\gamma_{G,k} F_k$ to be extracted for each frequency $\omega_k$:

$$\alpha_{ph}(T) = \alpha(T) - \alpha_{el}(T) = \frac{N_{Av} k_B \kappa_T}{V} \sum_k \gamma_{G,k} F_k \frac{x_k^2 e^{x_k}}{(e^{x_k}-1)^2} \qquad (13)$$

The spectral anharmonicity function, which we define as the PDOS weighted by the frequency-dependent Grüneisen parameter, $\gamma_{G,ph}(\omega) F(\omega)$, is represented in Figure 12a) and 12b) together with the PDOS obtained from the specific heat. The extracted values of frequency dependent Grüneisen parameter $\gamma_{G,ph}(\omega)$ are plotted in Figure 12c. For $Ba_8Si_{46}$ the modes below 10 meV are heavily weighted with $\gamma_{G,k}$ reaching values of up to 8.6. Higher energy modes are much less anharmonic with $\gamma_{G,k}$ values below 2. This is in reasonable agreement with what is found by Raman spectroscopy under pressure[49]. Anharmonicity does not play an important role in $Ba_{24}Si_{100}$: the spectral anharmonicity function displays similar peaks to the PDOS from the specific heat and only shows $\gamma_{G,k}$ values below 2.

Is this anharmonicity of the low-energy modes in $Ba_8Si_{46}$ in contradiction to our observation that the INS spectrum is temperature independent between 2 and 300 K? If we use the relation $\gamma_{G,ph} \equiv -\partial \ln \omega / \partial \ln V$ to estimate the shift of the phonon frequency from the measured expansion value $\Delta V/V = 7.2 \, 10^{-3}$ between 0 and 300 K, we find that the 7 meV mode is only shifted by 0.25 K. This is beyond the resolution of the INS experiment.

The pressure dependence of $T_c$ for $Ba_8Si_{46}$ is obtained from the Ehrenfest relation

$$\Delta\alpha = \frac{1}{3V} \frac{\Delta C}{T_c} \left(\frac{\partial T_c}{\partial p}\right)_T, \qquad (14)$$

where $\Delta\alpha$ and $\Delta C$ represent the discontinuities of $\alpha$ and $C$ at the second-order transition. The experimentally determined step $\Delta\alpha = -(1.1 \pm 0.2) \times 10^{-7}$ K$^{-1}$ (Figure 10) corresponds to $-1.0 \pm 0.2$ K/GPa for the initial pressure dependence of $T_c$. This value is in agreement with resistivity data under pressure[18]. Again assuming $\kappa_T^{-1} = 93$ GPa, we obtain the fractional volume dependence of the critical temperature $\partial \ln T_c / \partial \ln V = 11 \pm 2$. The fractional volume dependences of the critical temperature and Sommerfeld constant of $Ba_8Si_{46}$ are unusually large: 11 and -12.6 respectively.

## VII. Discussion and conclusions

The low-temperature specific heat of $Ba_8Si_{46}$ and $Ba_{24}Si_{100}$ has already been studied in previous publications[12,19]. Although small deviations are found, the superconducting parameters are more or less reproduced in the present work. However, a closer examination of the zero-field $Ba_8Si_{46}$ data reveals some excess specific heat which we could accurately fit using a 2-gap BCS model. Similar low energy excitations are also indicated in the STS spectrum of Ichimura *et al.*[34]. It would be instructive to see whether their data can also be fitted by a 2 gap model instead of the anisotropic gap scenario which the authors suggest. The low-temperature specific heat already provides two explanations for the higher $T_c$ in $Ba_8Si_{46}$: in the case of $MgB_2$ it has been shown that the presence of a second gap is crucial to explain its high $T_c$ of 40 K[30]. In $Ba_8Si_{46}$, such a second gap may help to enhance $T_c$ somewhat. However, the small weight of the second gap in the fit of only 10 % indicates that this effect is expected to be rather small. The second and certainly more important factor is the higher density of states at the Fermi level in $Ba_8Si_{46}$.

Regarding superconductors which are based on structures with ions in oversized crystalline cages, it appears that in many cases superconductivity is mediated by the phonon associated with the vibration of the encaged ion[2,3,4,30]. This was also proposed by Tse *et al.*[27] through *ab initio* calculations for the clathrate systems studied here. To test this scenario in the case of the clathrates, we first determined the PDOS by different methods; a deconvolution of the lattice specific heat, inelastic neutron and inelastic X-ray scattering. While in the present case the specific heat apparently provides a good approximation to the sample PDOS in absolute values (although with a limited energy resolution), the different scattering cross sections of the Ba and Si atoms influence the spectra from INS and IXS. Basically, two low-energy peaks may be extracted from the specific-heat PDOS for $Ba_8Si_{46}$ at 7 and 17 meV and an additional peak at 3 meV for $Ba_{24}Si_{100}$.

The INS and IXS experiments help us to interpret the nature of the peaks in the PDOS. The spectrum of $Ba_8Si_{46}$ below 25 meV is dominated by barium vibrations, which are responsible for the rapid upturn of the specific heat at low temperature. The peak in the 6 - 10 meV range can clearly be identified as due to barium vibrations. However, the second peak extracted from the deconvolution of the specific heat around 17 meV is formed by a distinct barium contribution at energies above 12 meV and hybridized Ba - Si modes at slightly higher energies. At even higher energies the spectrum is dominated by silicon modes. For $Ba_{24}Si_{100}$, the barium modes appear to be concentrated below 12 meV. An additional barium mode appears around 2.5 meV. The latter is responsible for the peak in the specific-heat PDOS around ~3meV and is related to the barium vibration in the additional large pseudo-cubic $Si_8$ cages. Due to this mode at particularly low energy the upturn in the low-temperature specific heat is even more pronounced than in $Ba_8Si_{46}$.

Using electrical resistivity as an experimental probe to obtain the electron-phonon coupling function from a deconvolution of the data into a set of Bloch-Grüneisen modes, helps us to investigate which of the peaks in the PDOS are related to a strong electron-phonon interaction and thus mainly responsible for the superconductivity. Our analysis shows that the principal coupling arises from the mode related to the 7 meV peak in the PDOS of $Ba_8Si_{46}$ associated with the $BaSi_{24}$ nanocage[42] and the 3 meV (2.5 from INS) mode in $Ba_{24}Si_{100}$. Assuming that $T_c$ scales linearly with phonon energy, this difference already explains an increase in $T_c$ by a factor 2-3 in $Ba_8Si_{46}$. The remaining difference may be partly provided by the larger density of states in $Ba_8Si_{46}$ and the presence of the second superconducting gap, as well as some differences in the host lattice cohesive energy[11]. The polycrystalline sample does not allow us to extract an absolute value for the electron-phonon coupling function. However, we know from the low-temperature specific heat (Section III) that the coupling strength is clearly larger in $Ba_8Si_{46}$. This may furthermore be indirectly reflected in the thermal expansion data: the low-energy modes below 10 meV in $Ba_8Si_{46}$ show a significant anharmonicity with large mode Grüneisen parameters of up to ~9. This anharmonicity of the low energy modes is clearly absent in $Ba_{24}Si_{100}$ with all mode Grüneisen parameters of ~2. Modes which show a strong electron-phonon coupling often show an enhancement of their mode Grüneisen parameter[2,3]. Hence, it may also indicate a smaller electron-phonon coupling in $Ba_{24}Si_{100}$ which explains the remaining difference in the $T_c$ values of both compounds.

A final question is whether the $T_c$ of $Ba_8Si_{46}$ could be raised further. Several factors are of importance in such materials based on ions enclosed in oversized crystalline cages: both the cage size and the mass of the enclosed ion have an influence on the phonon frequency which provides most of the electron-phonon coupling and is thus mainly responsible for $T_c$. However, at the same time the filling factor of the cages and hence the overlap of the electronic orbitals of the enclosed ion with the cage atoms are changed. The latter may be a crucial factor for the electron-phonon coupling strength. Applying pressure appears to be a good test to see whether $T_c$ could be raised further: since it reduces both the filling factor of the cage and the cage size. The smaller cage size increases the phonon frequency of the encaged ion and thus directly enhances $T_c$. Unfortunately, the cell parameter reduction due to pressure application also leads to an electronic band enlargement and consequently a reduction of the density of states at the Fermi level which seems to be the dominant factor controlling $T_c$, thus leading to the measured reduction of $T_c$ with pressure[18]. However, thermal expansivity at $T_c$ in combination with specific heat provides an answer without applying any pressure. The Ehrenfest relation which incorporates the jump at $T_c$ of both quantities (Eq. 14) tells us that the initial pressure dependence of $T_c$ is negative and $T_c$ may thus be enhanced by increasing the cell volume. Possibilities for an improvement of the $T_c$ values could then arise from the modification of the host-lattice properties: carbon appears then to be a good candidate for the host clathrate lattice as discussed in Ref.[29]. This will lead to the question of determining the appropriate guest atom for optimal electron-phonon coupling in the currently hypothetical carbon clathrates under the constraints of the reduced nanocage size.

## Acknowledgments

R.L. & Y.W. thank A. Junod for sharing his knowledge about calorimetry and its analysis with them. This work was partially supported by the National Science Foundation through the National Centre of Competence in Research "Materials with Novel Electronic Properties–MaNEP".

# Tables

**Table I.** Superconducting parameters of $Ba_8Si_{46}$ and $Ba_{24}Si_{100}$. The superconducting condensation energy $E_c$ and the thermodynamic critical field $H_c$ were obtained by numerically integrating the superconducting contribution to $C$ and $C/T$. In the case of $Ba_{24}Si_{100}$, the BCS fit shown in Figure 3b was used to extrapolate the data to zero temperature. The last two columns contain examples from the literature for comparison. See text for details.

| | $Ba_8Si_{46}$ | $Ba_{24}Si_{100}$ | Literature: $Ba_8Si_{46}$ | Literature: $Ba_{24}Si_{100}$ |
|---|---|---|---|---|
| $T_c$ [K] | 8.1 | 1.55 | 8.07[19] | 1.4[12], 1.55[11] |
| $\gamma_n$ [mJ/mole K$^2$] | 123.7 | 189.7 ($\gamma_{n\_res}$ corrected: 158.7) | 144[19] | 182[12] |
| $\Delta C/T_c$ [mJ/mole K$^2$] | 211.5 | 208.7 | 218.9[19] | 260.3[12] |
| $\Delta C/\gamma_n T_c$ | 1.71 | 1.1 ($\gamma_{n\_res}$ corrected: 1.43) | 1.52[19] | 1.43[12] ($\gamma_{n\_res}$ corrected) |
| $2\Delta_0/k_B T_c$ | $\Delta_{0\_1}$=1.3 $\Delta_{0\_2}$=4.4 | 3 ($\gamma_{n\_res}$ corrected 3.4) | 3.6[19] $\Delta_{0\_min}$: 1.3, $\Delta_{0\_max}$: 4.4[34] | 3.5[12] |
| $H_{c2}$ [T] | 5.75 | 0.35 | 5.5[33] | 0.35[11], 0.24[12] |
| $H_c$ [mT] | 23 | 2.8 | - | - |
| $E_c$ [J/mole] | 2.17 | 0.60 | - | - |
| $(1+\lambda_{ep})N(E_F)$ [eV cell]$^{-1}$ | 52.4 | 80.4 ($\gamma_{n\_res}$ corrected 63.9) | $(1+\lambda_{ep})$ x 31[19], $(1+\lambda_{ep})$ x 43[18], $(1+\lambda_{ep})$ x 38[25] | $(1+\lambda_{ep})$ x 60[38] |
| $(1+\lambda_{ep})N(E_F)$ [eV Ba-atom]$^{-1}$ | 6.6 | 3.35 ($\gamma_{n\_res}$ corrected 2.66) | $(1+\lambda_{ep})$ x 3.88[19], $(1+\lambda_{ep})$ x 5.38[18], $(1+\lambda_{ep})$ x 4.75[25] | $(1+\lambda_{ep})$ x 2.5[38] |

**Table II:** Scattering probabilities and proportionality factors for X-ray and neutrons in Ba and Si atoms.

| Element | $Z^2$ | $\sigma^n$ / **barns** | $M$ / **a.m.u.** | $Z^2/M$ | $\sigma^n/M$ |
|---------|-------|------------------------|------------------|---------|--------------|
| Si      | 196   | 2.2                    | 28.1             | 7.0     | **0.078**    |
| Ba      | **3136** | **3.4**             | **137.3**        | **22.8**| **0.025**    |

# Figure captions

**FIG. 1.** (Color online) Specific heat divided by temperature ($C/T$) of $Ba_8Si_{46}$ and $Ba_{24}Si_{100}$ at low temperatures showing the superconducting transitions at 8.1 K and 1.55 K, respectively. Inset: $C/T$ of $Ba_8Si_{46}$ and $Ba_{24}Si_{100}$ from 0 to 300 K (1 g.at = 1/54 mole and 1/124 mole for $Ba_8Si_{46}$ and $Ba_{24}Si_{100}$, respectively).

**FIG. 2.** (Color online): (a) $C/T$ and (b) resistivity of $Ba_8Si_{46}$ and (c) $C/T$ of $Ba_{24}Si_{100}$ at the superconducting transition in magnetic fields. The dotted lines in (a) and (c) are fits of the normal-state data (see text for details). The insets show the upper critical field line in a magnetic field vs. temperature phase diagram together with a standard WHH fit. In (a) data from resistivity is included where triangles mark $\rho$=0, squares the midpoint of the resistive transition and stars the onset of the transition. In (b) data from ac-susceptibility measurements[11] is included (filled squares).

**FIG. 3.** (Color online) Electronic contribution to the specific heat $C_e/T$ of $Ba_8Si_{46}$ (a) and $Ba_{24}Si_{100}$ (b) in zero field. In (a) fits according to a BCS model with $2\Delta_0/k_BT_c = 4$, a two gap model with $2\Delta_0/k_BT_c = 1.0$ (10 %) and $2\Delta_0/k_BT_c = 4.2$ (90 %) and an anisotropic s-wave model with $2\Delta_{min}/k_BT_c = 1.0$ and $2\Delta_{max}/k_BT_c = 4.2$ are added. In (b) circles represent the total electronic contribution together with the best possible BCS fit, while the square represents data that has been corrected for a possible small impurity residual $\gamma_n$ component (see text for details).

**FIG. 4.** (Color online) Lattice specific heat divided by temperature of $Ba_8Si_{46}$ (a) and $Ba_{24}Si_{100}$ (b) showing the decomposition into Einstein terms. The labels $k$ correspond to Einstein temperatures $\theta_{E,k} = 82$ K $\times 1.56^k$ – i.e. (from left to right) 9, 14, 22, 24, 53, 128, 200, 311, 486, 758 K.

**FIG. 5.** Phonon density of states $F(\omega)$ deconvolved from the specific heat for $Ba_8Si_{46}$ and $Ba_{24}Si_{100}$ on a logarithmic energy scale. Fits are performed with $\delta$-functions $F_k\delta(\omega-\omega_k)$ on a basis of Einstein frequencies $\omega_{k+1} = 1.56\omega_k$ (see Fig. 4). In order to reflect the spectral density, the $\delta$-functions of the PDOS are represented by a histogram of width $\Delta\omega_k \equiv 1.56^{1/2}\omega_k - \omega_k/1.56^{1/2}$ and height $F_k/\Delta\omega_k$.

**FIG. 6.** (Color online) Left: Generalized density of states $G(\omega)$ of $Ba_8Si_{46}$ as obtained by INS (red) and IXS (blue). Right, $G(\omega)$ of $Ba_{24}Si_{100}$ measured by INS on a linear temperature scale. All data sets are normalized to the same integral intensity in the energy range 2-85 meV. For presentation reasons IXS data have been divided by 3.25 corresponding to the effective intensity ratio of the barium and silicon inelastic responses. The PDOS (normalized) as obtained from the specific heat is added as a bar graph (see Figure 5).

**FIG. 7.** (Color online) Left: Generalized susceptibility of $Ba_8Si_{46}$ as obtained by INS at different temperatures. Right, generalized susceptibility of $Ba_{24}Si_{100}$ measured by INS. The grey shaded area indicates the energy range in which the data are obscured by the resolution function of the spectrometer.

**FIG. 8.** (Color online) Normal-state resistivity of $Ba_8Si_{46}$ (a) and $Ba_{24}Si_{100}$ (b) versus temperature. The dashed line represents the residual resistivity when superconductivity is quenched by a field of 5 T in (a). The labels $k$ correspond to Einstein temperatures $\theta_{E,k}$ =82 K x $1.56^k$ – i.e. (from left to right) 9, 14, 22, 24, 53, 128, 200, 311, 486, 758 K of the corresponding phonon modes.

**FIG. 9.** (Color online) Electron-phonon transport coupling function $\alpha_{tr}^2 F(\omega)$ (closed squares) deconvolved from the resistivity in comparison with the phonon density of states $F(\omega)$ deconvolved from the specific heat (histogram of rectangles) for $Ba_8Si_{46}$ (a) and $Ba_{24}Si_{100}$ (b). Fits are performed with δ-functions $(\alpha_{tr}^2 F)_k \delta(\omega - \omega_k)$ on a basis of Einstein frequencies $\omega_{k+1} = 1.56 \omega_k$. c) $\alpha_{tr}^2$ for both compounds.

**FIG. 10.** (Color online) Thermal expansivity $\alpha = 1/L \, dL/dT$ of $Ba_8Si_{46}$ and $Ba_{24}Si_{100}$. Inset: Thermal expansivity $\alpha/T$ versus $T^2$ at low temperature for $Ba_8Si_{46}$ and $Ba_{24}Si_{100}$. The straight lines are fits to extract the volume dependence of the Sommerfeld constant (see text for details).

**FIG. 11.** (Color online) Lattice thermal expansivity divided by temperature for $Ba_8Si_{46}$ (a) and $Ba_{24}Si_{100}$ (b) showing the decomposition into Einstein terms. The labels $k$ correspond to Einstein temperatures $\theta_{E,k}$ =82 K x $1.56^k$ – i.e. (from left to right) 9, 14, 22, 24, 53, 128, 200, 311, 486, 758 K.

**FIG. 12.** (Color online) Spectral anharmonicity function $\gamma_G(\omega) F(\omega) \equiv -(\partial \ln \omega / \partial \ln V) F(\omega)$ (closed squares) deconvolved from the thermal expansivity in comparison to the phonon density of states $F(\omega)$ deconvolved from the specific heat (histogram) for $Ba_8Si_{46}$ (a) and $Ba_{24}Si_{100}$ (b). Fits are performed with δ-functions $(\gamma_G F)_k \delta(\omega - \omega_k)$ on a basis of Einstein frequencies $\omega_{k+1} = 1.56 \omega_k$ (c) Frequency-dependence of the phonon Grüneisen parameter $\gamma_{G,ph}(\omega)$.

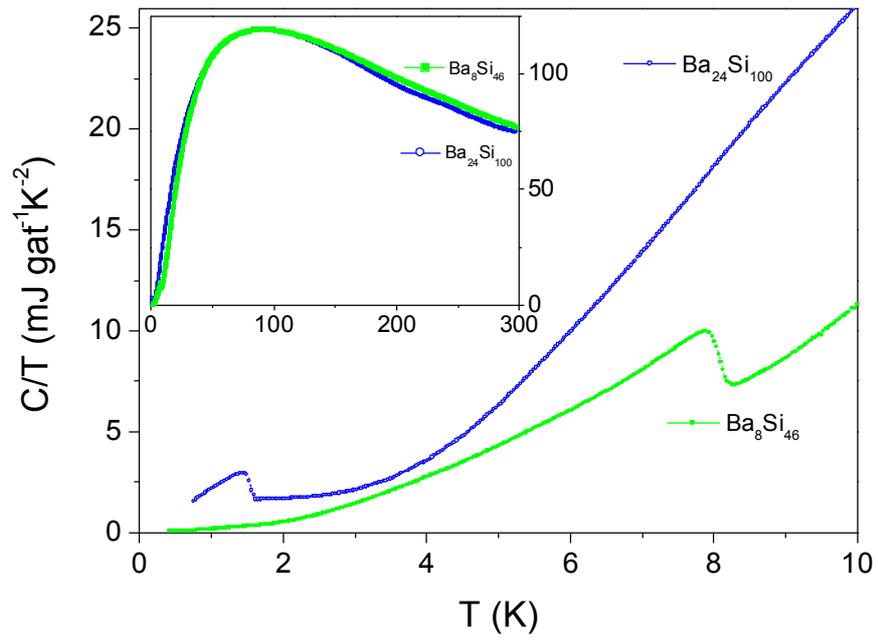

**FIG. 1**

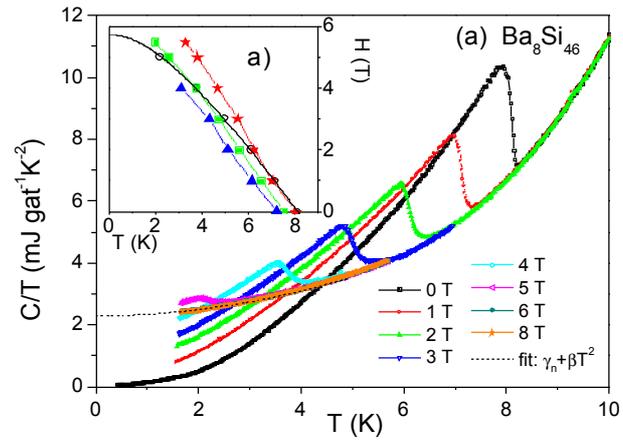

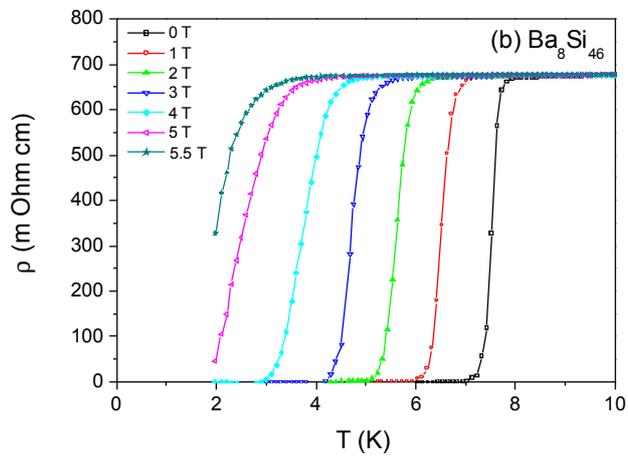

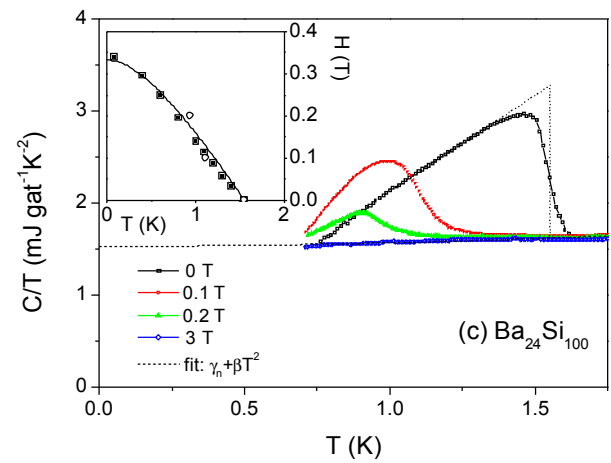

**FIG. 2a**
**FIG. 2b**
**FIG. 2c**

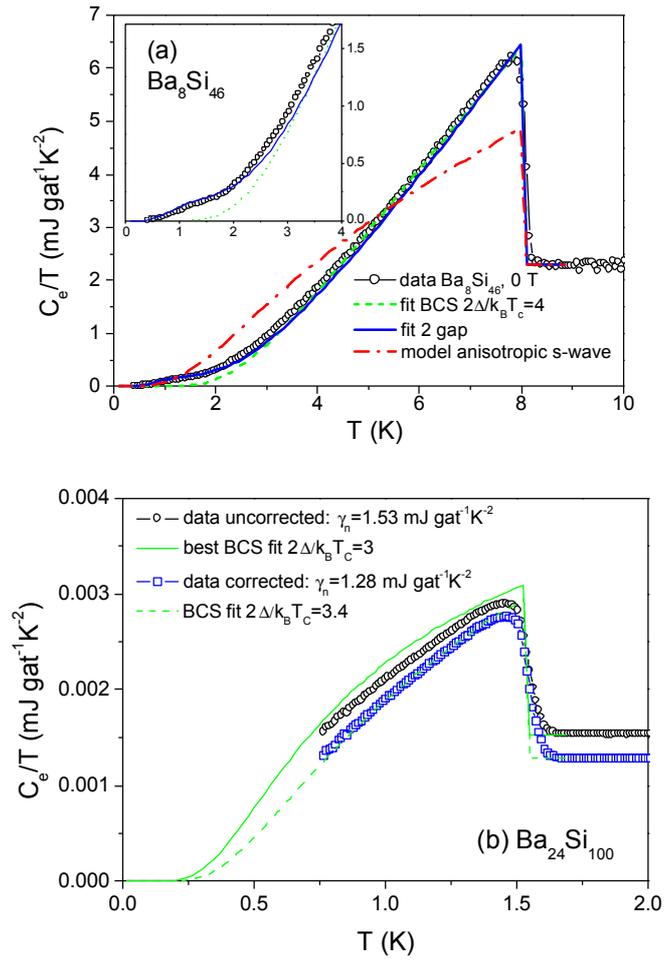

**FIG. 3a**
**FIG. 3b**

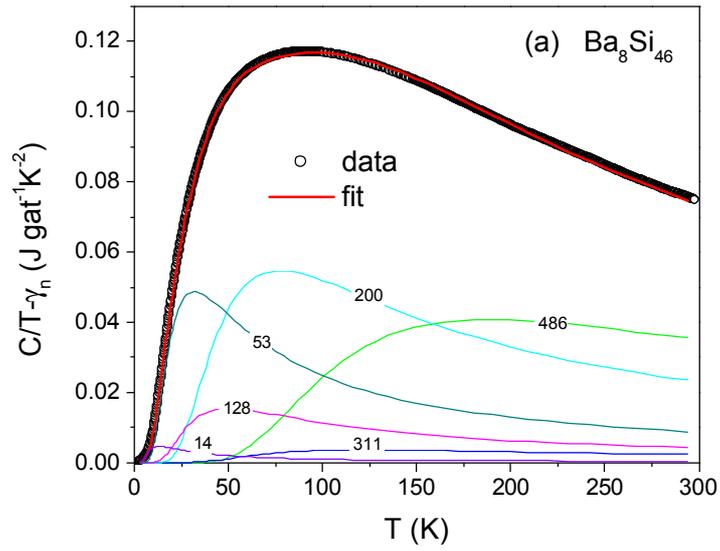

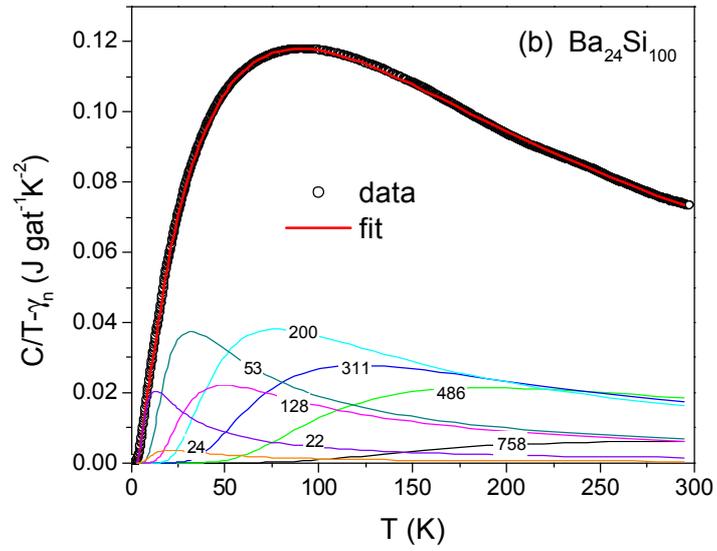

**FIG. 4a**
**FIG. 4b**

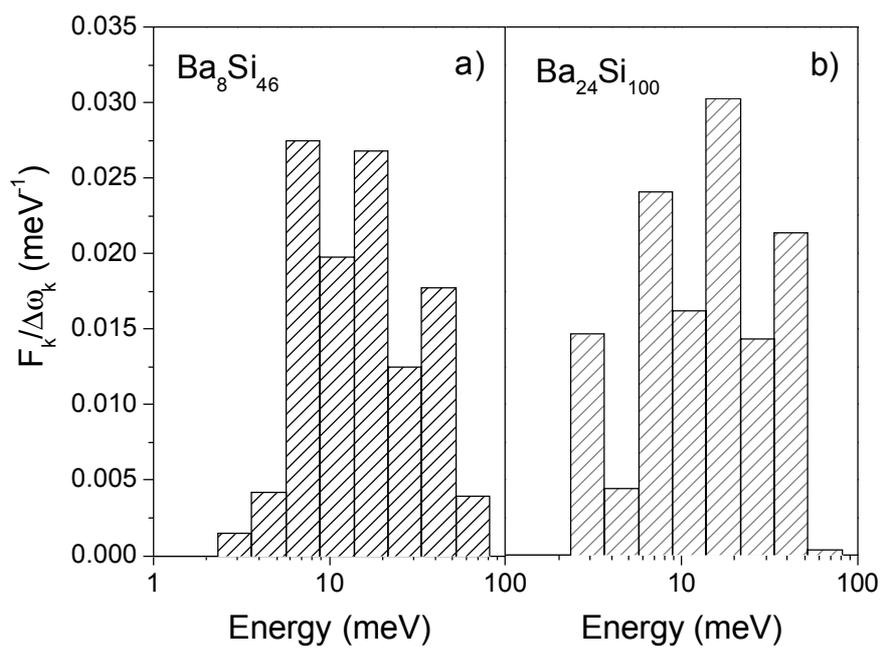

**FIG. 5**

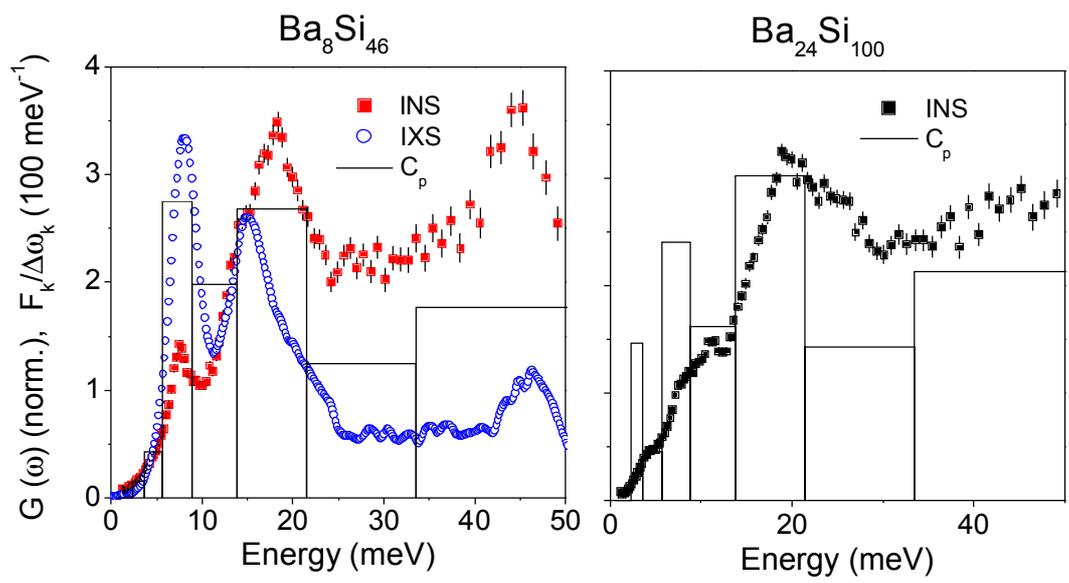

**FIG. 6**

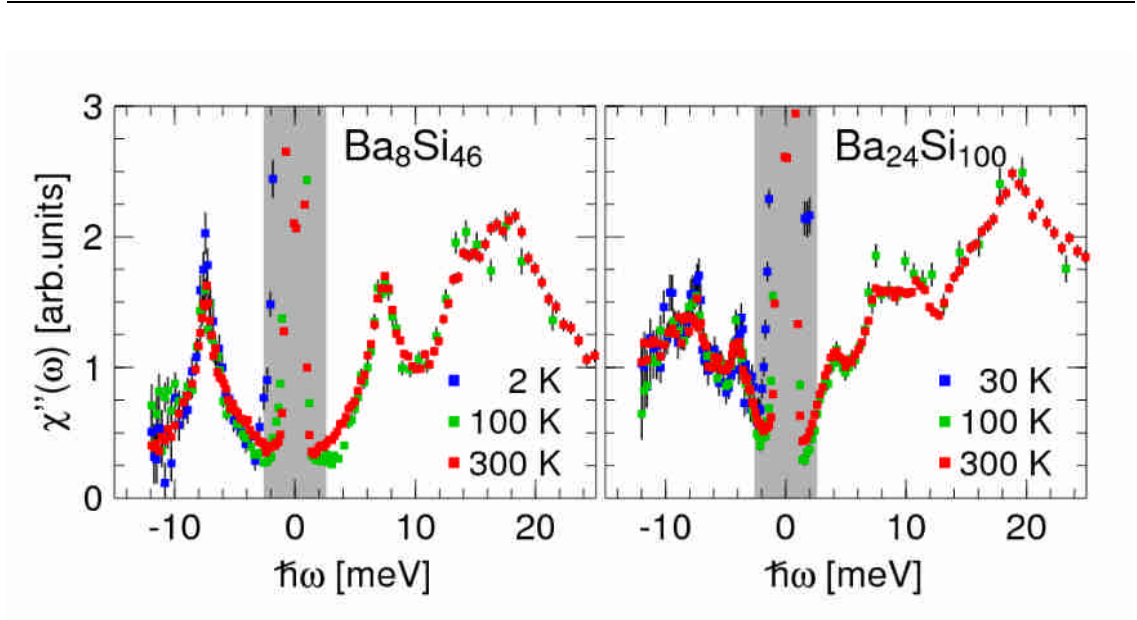

**FIG. 7**

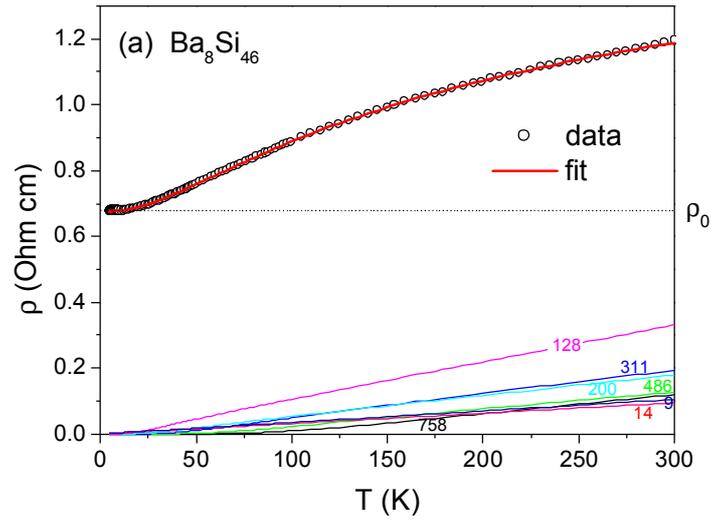

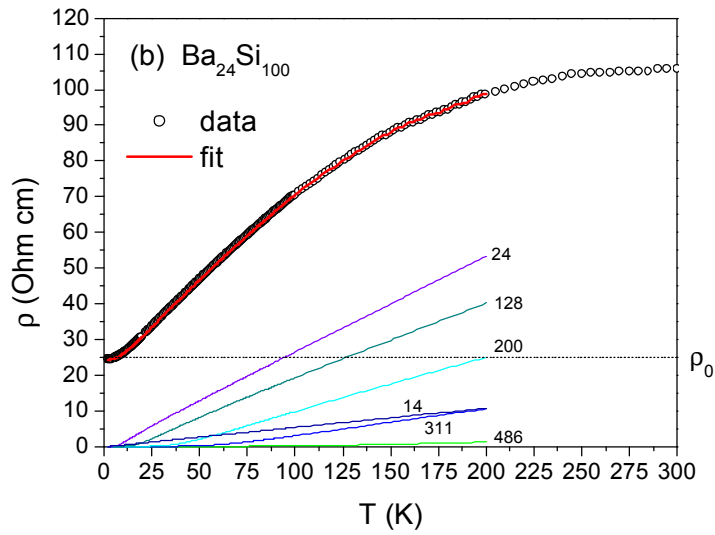

**FIG. 8a**

**FIG. 8b**

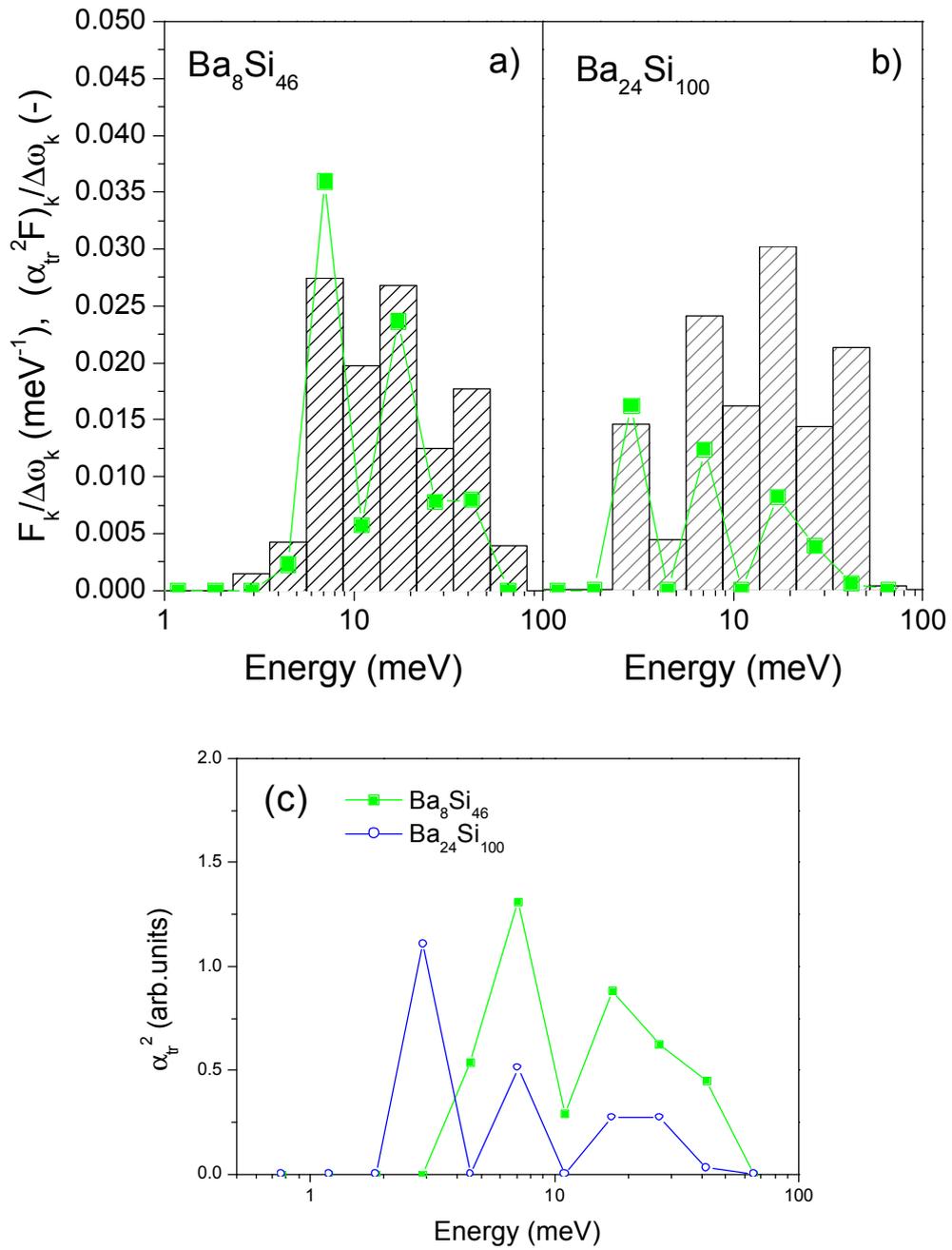

**FIG. 9a, FIG. 9b**

**FIG. 9c**

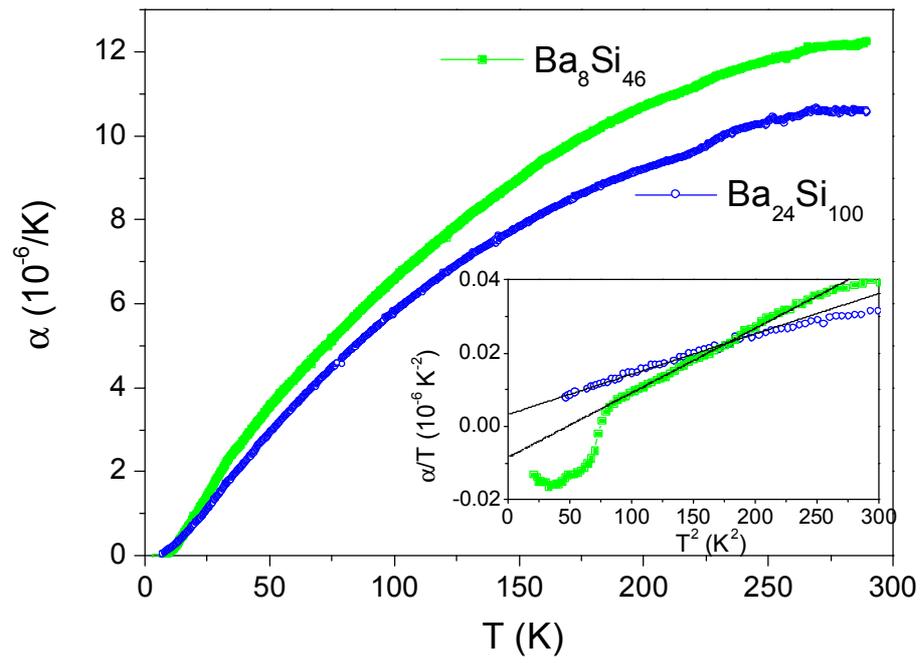

**FIG. 10**

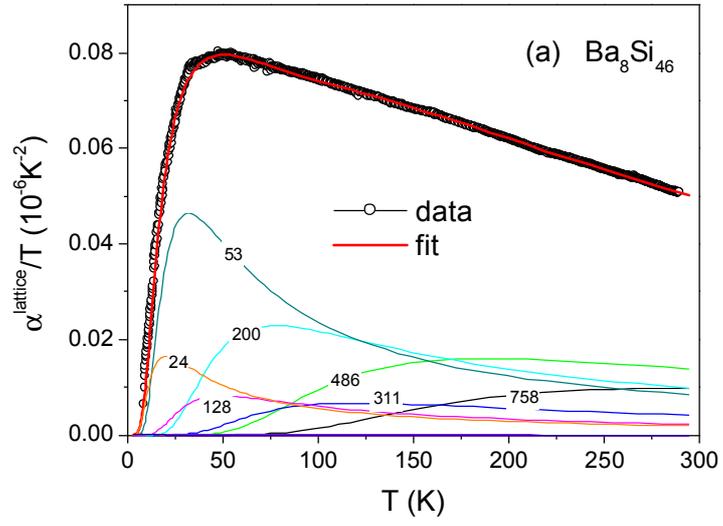

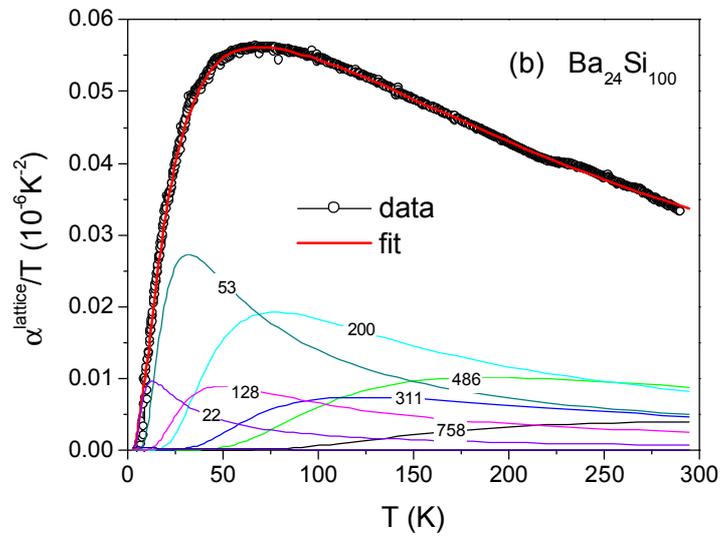

**FIG. 11a**

**FIG. 11b**

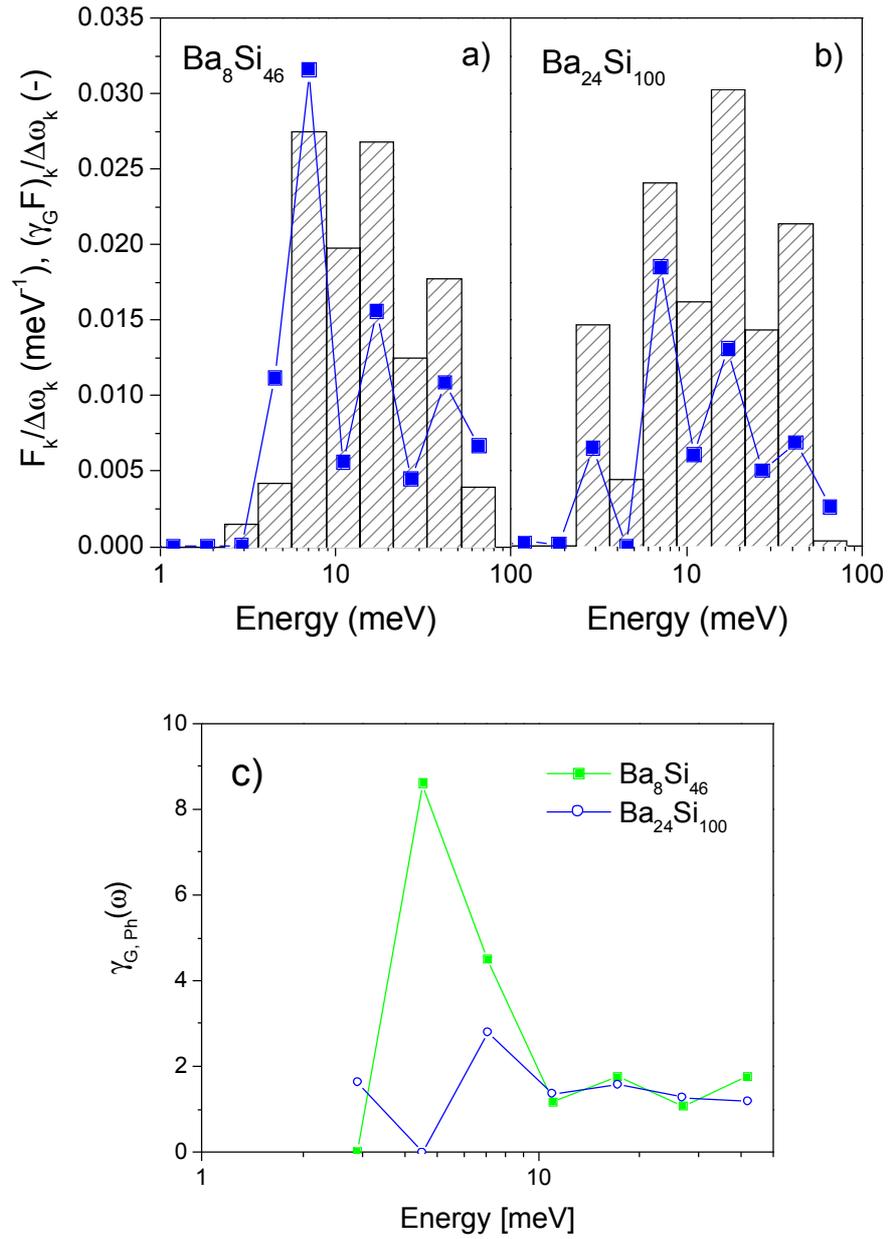

**FIG. 12a, FIG. 12b**
**FIG. 12c**